\documentclass[preprintnumbers,amsmath,amssymb,floatfix,10pt,prd,twocolumn,
superscriptaddress,nofootinbib]{revtex4}

\usepackage{graphicx}
\usepackage{dcolumn}
\usepackage{bm}
\usepackage{hyperref}
\usepackage{xcolor}
\usepackage{orcidlink}
\usepackage{latexsym}

\hypersetup{colorlinks=true, linkcolor=blue, citecolor=blue, urlcolor=blue}

\begin{document}

\title{Field Sources for Dark Matter Black Holes}

\author{G. Alencar \orcidlink{0000-0002-3020-4501}}
\email{geova@fisica.ufc.br}
\affiliation{Departamento de Física, Universidade Federal do Ceará, Fortaleza, Ceará, Brazil}

\author{C. R. Muniz \orcidlink{0000-0002-1266-2218}}
\email{celio.muniz@uece.br}
\affiliation{Universidade Estadual do Cear\'a, Faculdade de Educa\c c\~ao, Ci\^encias e Letras de Iguatu, 63500-000, Iguatu, CE, Brazil.}

\author{Francisco Tello-Ortiz \orcidlink{0000-0002-7104-5746}}
\email{francisco.tello@ufrontera.cl}
\affiliation{Departamento de Ciencias Físicas, Universidad de La Frontera, Casilla 54-D, 4811186 Temuco, Chile.}

\date{\today}

\begin{abstract}
We investigate the field-theoretic realization of regular black holes sourced by dark matter halo profiles within nonlinear electrodynamics (NED) minimally coupled to gravity. Starting from a static, spherically symmetric geometry determined by a halo density profile \(\rho(r)\), we reconstruct the associated mass function and derive the effective matter source supporting the spacetime. In the magnetic sector, the reconstruction is direct and yields a NED Lagrangian of the form \(L(F)=-\rho(r(F))\), while in the electric sector the theory is obtained parametrically through the field equations. We analyze the admissibility and consistency of the reconstructed models by studying regularity at the origin, asymptotic behavior, and the relevant energy conditions. The formalism is applied to representative halo profiles, including the Einasto, Dehnen, Burkert, and pseudo-isothermal families. For halo distributions with finite central density, the resulting geometries naturally exhibit de Sitter cores and asymptotically Schwarzschild behavior, providing a controlled and physically transparent link between dark matter halo phenomenology and regular black-hole spacetimes. Our results show that a broad class of halo profiles admits an effective NED completion, offering a unified geometric and field-theoretic interpretation of regular black holes sourced by dark matter halos.
\end{abstract}

\maketitle

\section{Introduction}
The nature of dark matter remains one of the central open problems in modern gravitational physics, astrophysics, and cosmology. Its existence is supported by a wide range of observational evidence, including galaxy rotation curves~\cite{Rubin1970}, galaxy cluster dynamics~\cite{Zwicky1933}, gravitational lensing observations such as the Bullet Cluster~\cite{Clowe2006}, and precision cosmological measurements~\cite{Planck2018}. For comprehensive reviews and modern perspectives on dark matter phenomenology see~\cite{Bertone2005,BullockBoylanKolchin2017,Salucci2019,deBlok2010}. At galactic scales, these effects are commonly modeled through phenomenological halo density profiles, which provide effective descriptions of the underlying mass distribution. Among the most widely used profiles are the Navarro–Frenk–White (NFW) model~\cite{Navarro1996}, the Einasto profile~\cite{Einasto1965}, the Dehnen family~\cite{Dehnen1993}, and cored distributions such as the Burkert profile~\cite{Burkert1995} and pseudo-isothermal halos~\cite{Begeman1991}. These models play a crucial role in connecting observational data with theoretical descriptions of dark matter.

On a different front, the problem of spacetime singularities in General Relativity has motivated the construction of regular black hole geometries, in which curvature invariants remain finite and the central singularity is replaced by a de Sitter core~\cite{Bardeen1968, Dymnikova1992, Hayward2006}. A particularly fruitful framework for generating such solutions is nonlinear electrodynamics (NED) minimally coupled to gravity, where nonlinear corrections to the electromagnetic sector modify the stress-energy tensor in a way that allows for singularity resolution. Seminal works by Ay\'on-Beato and Garc\'ia~\cite{AyonBeatoGarcia1998,AyonBeatoGarcia1999} demonstrated that regular black holes can arise from nonlinear electromagnetic sources, and subsequent studies have explored a wide variety of NED models and their properties~\cite{Bronnikov2001,Breton2004,Novello2000,Ansoldi2008,Kruglov2015}. The role of energy condition violations, in particular the localized breakdown of the strong energy condition, has been extensively analyzed as the key mechanism enabling regular geometries~\cite{MorenoSarbach2003,Ghosh2014,Zaslavskii2010}.
More generally, the relation between matter sources and spacetime geometry has been studied through reconstruction approaches, in which one attempts to infer the underlying field theory from a given metric. In the context of nonlinear electrodynamics, such inverse constructions have been developed in several works~\cite{Balart2014,FanWang2016,Toshmatov2017}, showing that different nonlinear Lagrangians can support the same effective geometry. This highlights the intrinsic non-uniqueness of the inverse problem and suggests that the geometry encodes only partial information about the underlying field dynamics.

A direct bridge between dark matter halo phenomenology and regular black hole physics has been recently established by Konoplya and Zhidenko~\cite{Konoplya:2025ect}, who showed that sufficiently dense halo profiles can act as effective sources of asymptotically flat, nonsingular black hole geometries. In this framework, the halo density profile $\rho(r)$ determines the mass function $m(r)$ and thus the spacetime geometry, provided an effective anisotropic equation of state is imposed. This result suggests that halo profiles, traditionally introduced as phenomenological descriptions of dark matter, may also encode effective matter sources capable of regularizing spacetime singularities.
Despite these developments, a fundamental question remains open. While halo-supported geometries can be constructed at the level of effective stress-energy tensors, the corresponding field-theoretic realization is not uniquely determined. In particular, it is not clear under which conditions a given halo density profile admits a consistent NED completion, nor how the properties of the halo translate into the ultraviolet and infrared structure of the underlying field theory. Addressing this issue requires a systematic reconstruction framework that connects halo phenomenology with nonlinear field dynamics.

In this work, we develop such a framework. Starting from a static, spherically symmetric geometry determined by a halo density profile, we reconstruct the corresponding stress-energy tensor and identify the NED Lagrangian that supports it. Our approach follows an inverse strategy, allowing us to interpret halo-supported regular black holes as effective solutions of NED minimally coupled to gravity.
Our analysis leads to several general results. First, we show that in the magnetic sector the reconstruction can be carried out explicitly, leading to a direct relation between the halo profile and the NED Lagrangian, $L(F) = -\rho(r(F))$. Second, in the electric sector the reconstruction is obtained in parametric form, making explicit the intrinsic degeneracy of the inverse problem. Third, and most importantly, we derive a set of general admissibility conditions on the halo profile that guarantee the existence of a physically consistent NED completion. These conditions ensure the regularity of the spacetime, the finiteness of curvature invariants, and the controlled behavior of the Lagrangian derivatives.
These results reveal a nontrivial correspondence between the asymptotic behavior of halo density profiles and the ultraviolet and infrared structure of NED theories. In particular, halo profiles with finite central density naturally generate an effective cosmological constant in the strong-field regime, leading to de Sitter cores, while their asymptotic decay determines the weak-field behavior of the reconstructed Lagrangian. This establishes a direct link between astrophysical halo phenomenology and the structure of nonlinear field theories.
We apply this framework to representative halo models, including the Einasto and Dehnen families, as well as cored profiles such as Burkert and pseudo-isothermal halos. In all cases, we show that the resulting geometries exhibit regular de Sitter cores and admit consistent NED interpretations. Our results suggest that halo-supported geometries may be understood as effective macroscopic manifestations of an underlying nonlinear field sector, providing a new perspective on the interplay between dark matter phenomenology and regular black hole physics.

The paper is organized as follows. In Sect.~\ref{sec2} we present the general reconstruction formalism. In Sect.~\ref{sec3} we analyze the consistency and non-uniqueness of the inverse problem. In Sect.~\ref{sec4} we derive general admissibility conditions for halo profiles. In Sects.~\ref{sec5} and \ref{sec6} we apply the method to specific halo models. In Sect. \ref{sec7} we discuss about the universal UV/IR correspondence between halo profiles and NED. Finally, in Sect.~\ref{sec8} we summarize our results and discuss possible extensions.

\section{General reconstruction of field sources from halo profiles}\label{sec2}

Consider a static spherically symmetric geometry

\begin{equation}\label{eq1}
ds^2
=
-f(r)dt^2
+\frac{dr^2}{f(r)}
+r^2 d\Omega^2 .
\end{equation}

Introducing the mass function,

\begin{equation}
f(r)=1-\frac{2m(r)}{r},
\end{equation}

the Einstein equations give

\begin{equation}\label{eq3}
\rho(r)=\frac{2m'(r)}{r^2},
\qquad
p_t(r)=-\frac{m''(r)}{r},
\end{equation}
and $p_{r}(r)=-\rho(r)$.
Therefore any halo density profile $\rho(r)$ uniquely determines the geometry through

\begin{equation}
m'(r)=\frac{\rho(r) r^2}{2}.
\end{equation}

Once the geometry is known, the corresponding field sources can be reconstructed.

\subsection{Magnetic NED}

For a magnetic monopole configuration

\begin{equation}\label{eq5}
F_{\theta\phi}=g\sin\theta ,
\end{equation}

the electromagnetic invariant becomes

\begin{equation}\label{eq6}
F=\frac{g^2}{2r^4}.
\end{equation}

In NED minimally coupled to gravity the stress tensor satisfies

\begin{equation}\label{eq7}
T^t{}_t=T^r{}_r=L(F),
\end{equation}

which implies

\begin{equation}\label{eq8}
L(r)=-\rho(r).
\end{equation}

Since $F(r)$ is known explicitly, the Lagrangian can be written as

\begin{equation}\label{eq9}
L(F)=-\rho\big(r(F)\big),
\qquad
r(F)=\left(\frac{g^2}{2F}\right)^{1/4}.
\end{equation}

Thus any halo density profile directly defines a NED theory supporting the corresponding geometry.

\subsection{Electric NED}

For a purely electric field

\begin{equation}\label{eq10}
F_{tr}=E(r),
\qquad
F=-\frac{E^2}{2},
\end{equation}

the stress tensor components read

\begin{equation}\label{eq11}
T^t{}_t=T^r{}_r=L-2F L_F,
\qquad
T^\theta{}_\theta=L .
\end{equation}

Combining these relations with the Einstein equations leads to the reconstruction condition

\begin{equation}\label{eq12}
F L_F
=
\frac{1}{2}
\left(
\frac{2m'(r)}{r^2}
-
\frac{m''(r)}{r}
\right).
\end{equation}

This relation determines the NED source in parametric form once the halo profile $\rho(r)$ is specified.

These relations show that dark matter halo geometries admit a natural field-theoretic interpretation. In particular, any density profile $\rho(r)$ defines a corresponding NED theory minimally coupled to gravity that supports the same spacetime geometry.

\subsection{Energy conditions and physical viability}

A crucial aspect in the interpretation of halo-supported geometries as field-theoretic configurations is the analysis of the energy conditions \cite{Curiel2017}. 

\paragraph{Weak Energy Condition (WEC).}
The WEC requires
\begin{equation}
\rho \geq 0, \qquad \rho + p_r \geq 0, \qquad \rho + p_t \geq 0.
\end{equation}
The first condition is automatically satisfied for physically meaningful halo profiles. The second condition yields $\rho + p_r = 0$, which is marginally satisfied. The third condition imposes
\begin{equation}
\rho - \frac{m''(r)}{r} \geq 0.
\end{equation}

\paragraph{Null Energy Condition (NEC).}
The NEC is equivalent to the condition $\rho+p_t\geq 0$, and therefore is controlled by the concavity of the mass function.

\paragraph{Dominant Energy Condition (DEC).}
The DEC requires
\begin{equation}
\rho \geq |p_r|, \qquad \rho \geq |p_t|.
\end{equation}
Since $p_r=-\rho$, the radial condition is saturated, while the tangential one gives
\begin{equation}
\rho \geq \left|\frac{m''(r)}{r}\right|.
\end{equation}

For halo profiles with finite central density, one typically finds
\begin{equation}
m(r)\sim r^3 \qquad \text{as} \qquad r\to 0,
\end{equation}
which implies regular behavior of all stress-energy components. In particular, near the origin,
\begin{equation}
\rho \approx \rho_0, \qquad p_t \approx -\rho_0,
\end{equation}
indicating an effective de Sitter core. These observations support the physical viability of the reconstructed matter sector at least in the central region.

\section{Consistency and Uniqueness of the Reconstructed NED Sector}\label{sec3}

The reconstruction procedure presented in Sec.~\ref{sec2} establishes a direct map between a given halo density profile $\rho(r)$ and a NED Lagrangian through the relation $L(F) = -\rho(r(F))$ in the magnetic sector, and via a parametric relation in the electric case. It raises a fundamental question regarding the uniqueness and physical consistency of the resulting field theory.

\subsection{Non-uniqueness of the inverse problem}

The reconstruction of a matter Lagrangian from a prescribed geometry constitutes an inverse problem in gravitational physics. In general, such problems do not admit a unique solution. In the present context, the ambiguity arises from the fact that the stress-energy tensor only fixes a subset of the possible functional degrees of freedom of the underlying field theory.

In the magnetic sector, the relation \eqref{eq9}
appears to define a unique Lagrangian. However, this identification implicitly assumes a minimal coupling and a specific identification between the invariant $F$ and the radial coordinate. More generally, one may consider alternative field parametrizations or Legendre-dual formulations of NED, leading to inequivalent Lagrangians that generate the same energy-momentum tensor \cite{AyonBeatoGarcia1998,AyonBeatoGarcia1999,Bronnikov2001,Plebanski1970}. On the other hand, in the electric sector, the situation is even more transparent. The reconstruction equation
\eqref{eq12} defines the theory only implicitly, allowing for different functional forms of $L(F)$ consistent with the same parametric curve. This explicitly demonstrates that the mapping from geometry to field theory is not one-to-one.

This degeneracy implies the existence of an equivalence class of NED models supporting a given halo geometry. From a physical perspective, this suggests that the geometry encodes only effective information about the underlying field theory, rather than uniquely determining its microscopic structure.

\subsection{Consistency conditions for the reconstructed theory}

To reduce this degeneracy, it is necessary to impose additional physical conditions on the reconstructed Lagrangian. In NED, these typically include:

\begin{itemize}
\item \textbf{Maxwell limit:} The theory should reduce to standard electrodynamics in the weak-field regime,
\begin{equation}
L(F) \sim F, \quad F \to 0.
\end{equation}

\item \textbf{Positivity and causality:} The absence of superluminal modes and instabilities requires
\begin{equation}
L_F > 0, \quad L_{FF} \geq 0.
\end{equation}

\item \textbf{Energy conditions:} The effective energy density should remain positive and physically meaningful throughout the spacetime.
\end{itemize}

In the present framework, these conditions impose nontrivial constraints on the allowed halo profiles. For instance, profiles leading to $L(F \to 0) \neq F$ correspond to strongly non-Maxwellian theories in the infrared regime, while profiles with rapidly varying curvature may violate stability conditions through negative $L_F$.

The existence of multiple field-theoretic realizations for the same halo-supported geometry suggests that the reconstructed NED should be interpreted as an effective description rather than a fundamental theory. In this sense, dark matter halo profiles may encode the macroscopic imprint of an underlying nonlinear field sector, whose precise form cannot be uniquely inferred from geometry alone.

This perspective places halo-supported regular black holes within the broader context of effective field theories in gravity, where different microscopic models may lead to indistinguishable macroscopic geometries.

\section{Admissibility Conditions for Halo Profiles and Physical NED Completions}\label{sec4}

The existence of a NED completion for a halo-supported geometry does not by itself guarantee the physical consistency of the resulting theory. Beyond the mere reconstruction of the matter sector, additional requirements must be imposed in order to ensure the regularity of the spacetime, the viability of the effective energy distribution, and the consistency of the field dynamics in both the weak- and strong-field regimes.

This naturally raises the question of whether such conditions can be formulated directly at the level of the halo profile itself, independently of the detailed reconstruction procedure. In other words, one may ask which properties of a density profile $\rho(r)$ are sufficient to guarantee that the associated geometry admits a physically acceptable NED interpretation.

In this section we derive a set of general admissibility conditions for halo-supported geometries. These criteria provide a direct characterization of the classes of density profiles capable of generating regular NED configurations.

\subsection{General derivatives of the reconstructed Lagrangian}

In the magnetic sector, the invariant is
given by Eq.~\eqref{eq6},
so that
\begin{equation}
\frac{dF}{dr}=-\frac{2g^2}{r^5}=-\frac{4F}{r}.
\label{eq:dFdr}
\end{equation}
Using Eq.~\eqref{eq8}, one finds
\begin{equation}
L_F=\frac{dL/dr}{dF/dr}
=\frac{-\rho'(r)}{-4F/r}
=\frac{r\,\rho'(r)}{4F}.
\label{eq:LFgeneral}
\end{equation}
Then, 
\begin{equation}
L_F=\frac{r^5}{2g^2}\,\rho'(r).
\label{eq:LFgeneral2}
\end{equation}
Differentiating once more,
\begin{equation}
L_{FF}=\frac{dL_F/dr}{dF/dr},
\label{eq:LFFdef}
\end{equation}
a direct computation yields
\begin{equation}
L_{FF}
=
-\frac{r^9}{8g^4}\left(5\rho'(r)+r\rho''(r)\right).
\label{eq:LFFgeneral}
\end{equation}

Therefore, the consistency properties of the reconstructed NED theory are entirely controlled by the first and second derivatives of the halo density profile.

\subsection{Regularity criterion}

A central requirement is that the geometry be regular at the origin. So from the first expression in Eq.~\eqref{eq3}
\begin{equation}
m'(r)=\frac{1}{2}\rho(r)r^2.
\label{eq:mprime}
\end{equation}
If the density remains finite near the center,
\begin{equation}
\rho(r)=\rho_0+O(r^\alpha), \qquad \alpha>0,
\label{eq:rhoreg}
\end{equation}
then
\begin{equation}
m(r)\sim \frac{\rho_0}{6}r^3,
\label{eq:mregular}
\end{equation}
and hence
\begin{equation}
f(r)=1-\frac{2m(r)}{r}\sim 1-\frac{\rho_0}{3}r^2.
\label{eq:desittercore}
\end{equation}
This is precisely the de Sitter-core behavior associated with regular black hole geometries \cite{Dymnikova1992}. Thus, finite central density is a sufficient condition for curvature regularity.

\subsection{Admissibility theorem}

We may now formulate the following result.

\medskip

\noindent
\textbf{Theorem.}
\textit{
Let $\rho(r)$ be a smooth, non-negative halo density profile defining a static spherically symmetric geometry through Eq.~\eqref{eq:mprime}. Assume that:
\begin{enumerate}
\item $\rho(r)$ is finite at the origin, with
\begin{equation}
\rho(r)=\rho_0+O(r^\alpha), \qquad \rho_0>0, \quad \alpha>0;
\end{equation}
\item $\rho(r)$ is monotonically decreasing for $r>0$,
\begin{equation}
\rho'(r)\leq 0;
\end{equation}
\item the combination
\begin{equation}
5\rho'(r)+r\rho''(r)\leq 0
\end{equation}
holds for all $r>0$;
\item asymptotically, the density decays sufficiently fast so that
\begin{equation}
\rho(r)\to 0, \qquad m(r)\to M<\infty
\end{equation}
as $r\to\infty$.
\end{enumerate}
Then the corresponding magnetic reconstruction defines a nonlinear electrodynamics completion with the following properties:
\begin{enumerate}
\item the spacetime possesses a regular de Sitter core at the center;
\item the Lagrangian remains finite in the strong-field regime,
\begin{equation}
L(F\to\infty)\to -\rho_0;
\end{equation}
\item the first derivative satisfies
\begin{equation}
L_F\leq 0;
\end{equation}
\item the second derivative satisfies
\begin{equation}
L_{FF}\geq 0.
\end{equation}
\end{enumerate}
Therefore, $\rho(r)$ admits a well-defined effective nonlinear electrodynamics completion with regular ultraviolet behavior and controlled convexity properties.
}

\medskip

\noindent
\textbf{Proof.}
Condition (1) implies Eq.~\eqref{eq:mregular}, which in turn gives the de Sitter-core behavior \eqref{eq:desittercore}; hence the spacetime is regular at the center. Condition (4) guarantees that the mass function tends to a finite constant and the geometry becomes asymptotically Schwarzschild.

From Eq.~\eqref{eq9}, one directly obtains
\begin{equation}
L(F\to\infty)= -\rho(0)=-\rho_0,
\end{equation}
which proves the finiteness of the strong-field limit. Next, using Eq.~\eqref{eq:LFgeneral2}, condition (2) implies
\begin{equation}
L_F=\frac{r^5}{2g^2}\rho'(r)\leq 0.
\end{equation}
Finally, Eq.~\eqref{eq:LFFgeneral} together with condition (3) gives
\begin{equation}
L_{FF}
=
-\frac{r^9}{8g^4}\left(5\rho'(r)+r\rho''(r)\right)\geq 0.
\end{equation}
This completes the proof.
\hfill $\square$

The theorem should be interpreted as an admissibility criterion rather than as a uniqueness statement. It provides sufficient conditions under which a halo profile defines a physically controlled effective NED completion in the magnetic sector. In particular, it shows that the qualitative properties commonly attributed to realistic cored halos --- finite central density, monotonic decrease, and smooth concavity --- are precisely the properties needed to obtain a regular and well-behaved nonlinear field theory.

An important consequence is that not all halo profiles are equally acceptable from the field-theoretic point of view. Profiles that are too steep, non-monotonic, or singular at the origin may still define a formal geometry, but they need not admit a NED completion with controlled derivative structure. The theorem therefore provides a physically motivated selection principle in the space of halo-supported regular black holes.

\subsection{Corollary: power-law tails and weak-field structure}

If, in addition, the halo density satisfies
\begin{equation}
\rho(r)\sim r^{-n}, \qquad r\to\infty,
\end{equation}
then, since $r\sim F^{-1/4}$, the reconstructed Lagrangian behaves as
\begin{equation}
L(F)\sim -F^{n/4}, \qquad F\to 0.
\end{equation}
Thus the asymptotic decay of the halo profile determines the infrared structure of the effective NED theory. In particular, only the special case $n=4$ would formally mimic a Maxwell-like scaling $L\sim F$ in the weak-field regime. More general halo tails lead instead to intrinsically non-Maxwellian infrared completions.

\section{Einasto halo solutions and field reconstruction}\label{sec5}

In Ref.~\cite{Konoplya:2025ect} several dark matter halo profiles were shown to generate regular black hole geometries. In particular, the authors considered the Einasto family

\begin{equation}
\rho(r)=\rho_0
\exp\!\left[
-\left(\frac{r}{h}\right)^{1/n}
\right],
\qquad n>0 .
\end{equation}

Two representative cases analyzed in \cite{Konoplya:2025ect} are $n=\tfrac12$ and $n=1$. Below we reconstruct the corresponding NED sources.

\subsection{Gaussian case $n=\tfrac12$}

For $n=\tfrac12$ the density profile becomes Gaussian,

\begin{equation}\label{einastoprofile1}
\rho(r)=\rho_0 e^{-r^2/h^2}.
\end{equation}
The central density $\rho_0$ is not an independent parameter but is determined by the total mass $M$ of the system. By integrating the density profile (\ref{einastoprofile1}) over the entire volume 
\begin{equation}
    M = \int_0^\infty 4\pi r^2 \rho(r) dr,
\end{equation}
we obtain the normalization condition for the Gaussian case
\begin{equation}
\rho_0 = \frac{M}{\pi \sqrt{\pi} h^3}.
\end{equation}
This relation ensures that the mass function $m(r)$ asymptotically approaches the total mass $M$ as $r \to \infty$, consistent with the Schwarzschild behavior at large distances. Furthermore, it highlights that for a fixed total mass, the central density scales as $h^{-3}$, implying that more compact halos (smaller $h$) require significantly higher energy densities to support the same gravitational mass.

The corresponding geometry obtained in \cite{Konoplya:2025ect} is

\begin{equation}\label{f(r)erf}
f(r)
=
1-\frac{2m(r)}{r}
=
1-\frac{2M}{r}\,
\mathrm{erf}\!\left(\frac{r}{h}\right)
+
\frac{4M}{\sqrt{\pi}h}
e^{-r^2/h^2}.
\end{equation}

\subsubsection{Magnetic sector}

For a magnetic monopole given by expressions \eqref{eq5}-\eqref{eq6}, the Eq.~\eqref{eq7} holds. This directly yields Eq.~\eqref{eq8}. So, using the right expression in Eq.~\eqref{eq9} the nonlinear electrodynamics Lagrangian becomes

\begin{equation}
L(F)=
-\rho_0
\exp\!\left[
-\frac{1}{h^2}
\left(\frac{g^2}{2F}\right)^{1/2}
\right].
\end{equation}

\subsubsection{Electric sector}

For a purely electric field, the electric NED source is determined parametrically from expressions \eqref{eq10}-\eqref{eq12}. Furthermore, the electric field can be explicitly determined by applying the generalized Gauss's law $\nabla_\mu (L_F F^{\mu\nu}) = 0$. For a static and spherically symmetric metric, this leads to the generalized Gauss law $r^2 L_F E = q$, where 
$q$ is the integration constant associated with the electric charge. Given the mass function $m(r)$ presented in Eq. (\ref{f(r)erf}) for this profile , we can derive the tangential pressure from Eq.~\eqref{eq3}, which together with the density $\rho(r) = \rho_0 e^{-r^2/h^2}$, leads to the following expression for the electric field
\begin{equation}
E(r) = -\frac{\rho_0}{q h^2} r^4 e^{-r^2/h^2}.
\end{equation}
This result highlights a particularly smooth behavior at the origin, where the electric field vanishes as $r^4$. This suppression is stronger than in the $n=1$ case, as we will see, further reinforcing the regularity of the spacetime geometry and the formation of a de Sitter core. 
\begin{figure}[h!]
    \centering
    \includegraphics[width=0.9\linewidth]{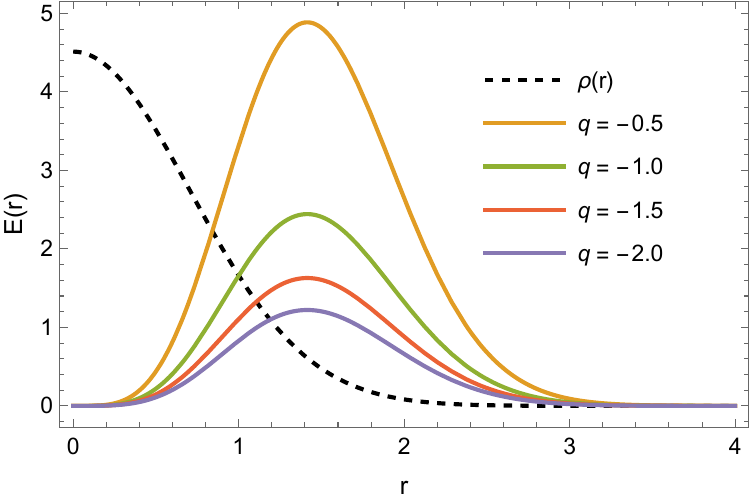}
    \includegraphics[width=0.9\linewidth]{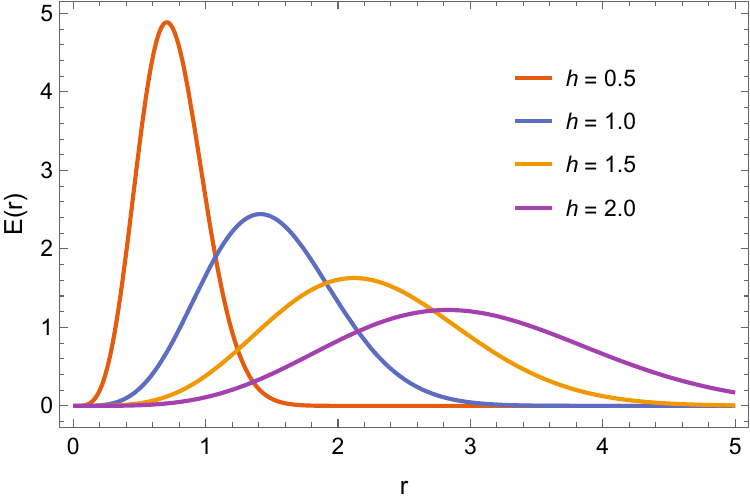}
\caption{Reconstructed electric field $E(r)$ for the $n=1/2$ Einasto profile. Top panel: Electric field behavior for varying charge $q$ with fixed $M=1$ and $h=1$; the inverse relation between $E(r)$ and $|q|$ highlights the non-linear nature of the reconstruction. Bottom panel: Electric field evolution for different scale parameters $h$ with $M=1$ and $q=-1$.}\label{elefieldgaussian}
\end{figure}

The reconstructed electric field $E(r)$ for the Gaussian Einasto profile ($n=1/2$) is displayed in Fig. \ref{elefieldgaussian}. The top panel illustrates the field behavior for varying the charge $q$ while keeping the total mass $M=1$ and the scale parameter $h=1$ fixed. As observed, there is an inverse relationship between the field strength and $q$, which highlights the intrinsically nonlinear nature of the reconstruction process. In the bottom panel, we examine the evolution of $E(r)$ for different values of the scale parameter $h$, with fixed $M=1$ and $q=-1$. The results show that as $h$ increases, the field peak undergoes both a spatial displacement and a gradual attenuation. This behavior demonstrates how the NED sector adjusts to accommodate the dark matter distribution while ensuring the formation of a regular de Sitter core. The strong suppression at the center, combined with the super-exponential decay at large distances, reflects the highly localized nature of the Gaussian halo source and reinforces the overall regularity of the spacetime geometry.

The electric sector associated with the Gaussian Einasto profile provides a robust mechanism for the black hole regularization. The nonlinear dynamics of the electromagnetic field leads to a stress-energy tensor where the radial pressure satisfies $p_r = -\rho$ globally. Consequently, the Strong Energy Condition (SEC) simplifies to $\rho + p_r + 2p_t = 2p_t$. As derived from the geometry, the tangential pressure at the origin behaves as $p_t \approx -\rho_0$, resulting in a local violation of the SEC ($2p_t \approx -2\rho_0 < 0$). This specific violation is the fundamental driver behind the formation of the de Sitter core, providing the repulsive gravitational effect necessary to prevent the central singularity \cite{Zaslavskii2010}. Furthermore, this framework is physically consistent with the evasion of Penrose’s singularity theorems, as the localized violation of the energy conditions within the scale radius $h$ allows for a non-singular spacetime despite the presence of a trapped surface. Unlike models with global exotic matter, the Gaussian decay ensures that these effects are strictly confined, leading to a rapid restoration of standard energy conditions in the weak-field regime.
\begin{figure}[h!]
    \centering
    \includegraphics[width=0.9\linewidth]{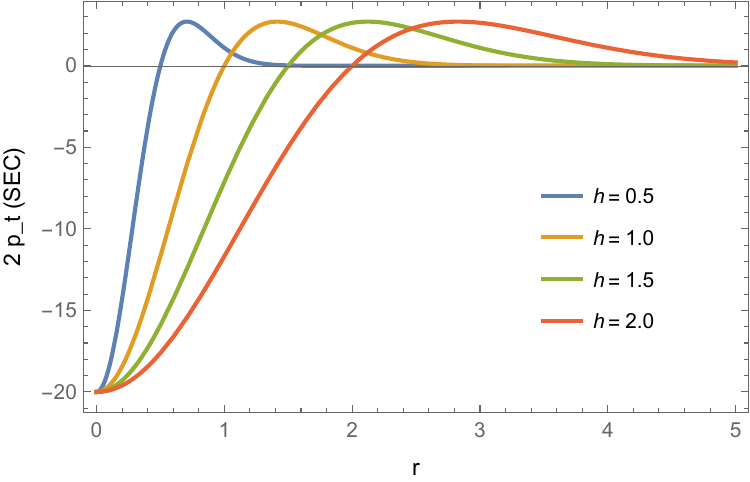}
    \includegraphics[width=0.9\linewidth]{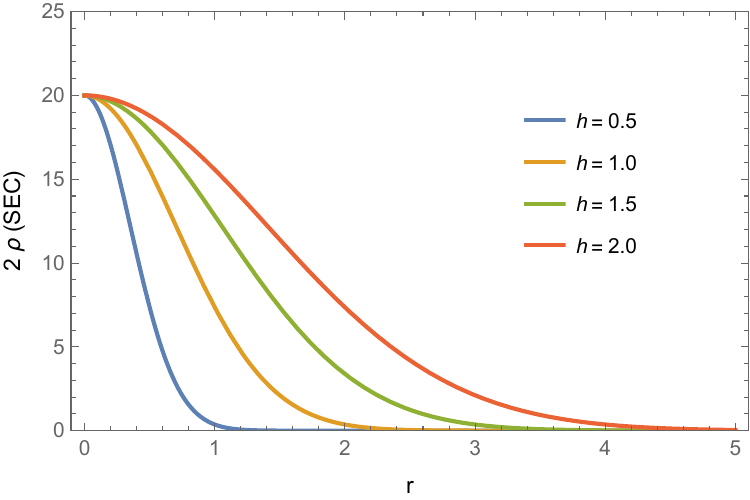}
\caption{Analysis of the Strong Energy Condition (SEC) for the electric sector of a regular black hole with an Einasto dark matter halo ($n=1/2$) and $\rho_0=10$. The top panel shows the tangential pressure combination $2p_t$, indicating SEC violation near the origin. The bottom panel displays the radial term $2\rho$, serving as a reference for the energy density of the halo. Curves represent different scale parameters $h$.}
\label{SEC-Gaussian-electric}
\end{figure}

Figure \ref{SEC-Gaussian-electric} confirms that the electric sector drives the SEC violation. The top panel shows that $2p_t$ attains $-2\rho_0$ at $r=0$, providing the repulsive tension needed for the de Sitter core. As $r$ increases, the pressure becomes positive before decaying, with smaller $h$ values yielding more localized violation zones. Crucially, the radial sector saturates the null energy condition ($\rho + p_r = 0$) globally, as established by the reconstruction framework. This saturation, combined with the standard monotonic decay of $2\rho$ in the bottom panel, ensures that exotic effects are restricted to the high-curvature regime. Thus, the Gaussian Einasto profile enables an efficient regularization that vanishes rapidly in the weak-field limit.

\subsection{Exponential case $n=1$}

For $n=1$ the Einasto profile reduces to

\begin{equation}
\rho(r)=\rho_0 e^{-r/h}.
\end{equation}
Similarly to the Gaussian case, the normalization for the exponential profile ($n=1$) requires the total mass $M$ to be related to the central density by $M = 8\pi \rho_0 h^3$. This ensures that the mass function correctly describes a system with finite total energy. In this configuration, the electric field (derived via the reconstruction procedure) exhibits a broader spatial distribution compared to the $n=1/2$ case, consistent with the slower decay of the Einasto profile for higher values of $n$.

The metric obtained in \cite{Konoplya:2025ect} is

\begin{equation}
f(r)=1-\frac{2m(r)}{r}
=
1-\frac{2M}{r}
+\frac{M}{2h}\,
\frac{2h^2+2hr+r^2}{h^2 r}
e^{-r/h}.
\end{equation}

The mass function is therefore

\begin{equation}
m(r)=
M-\frac{M}{4h^3}(2h^2+2hr+r^2)e^{-r/h}.
\end{equation}

\subsubsection{Magnetic sector}

For a magnetic monopole the invariant is again
given by expressions \eqref{eq9}. So that

\begin{equation}
L(F)=
-\rho_0
\exp\!\left[
-\frac{1}{h}
\left(\frac{g^2}{2F}\right)^{1/4}
\right].
\end{equation}

\subsubsection{Electric sector}

In the electric configuration the reconstruction equation \eqref{eq12} becomes
\begin{equation}
F L_F =
\frac{M}{8h^5}\,
r(r+4h)e^{-r/h},
\end{equation}
where we used the mass function found above.
Therefore the electric NED source is again determined parametrically.

To find the explicit form of the electric field $E(r)$, we utilize the non-linear 
Maxwell equations $\nabla_\mu (L_F F^{\mu\nu}) = 0$ once more. By substituting $L_F = q/(r^2 E)$ and the electric invariant $F = -E^2/2$ into the reconstruction condition, Eq. (\ref{eq12}), we obtain
\begin{equation}\label{elefieldreconst}
-\frac{qE}{2r^2} = \frac{1}{2}\left(\frac{2m'(r)}{r^2}-\frac{m''(r)}{r}\right).
\end{equation}
For the specific case of the $n=1$ Einasto profile, using the result from Eq. (33), 
 the electric field is uniquely determined as
\begin{equation}
E(r) = -\frac{M}{4qh^5}r^3(r+4h)e^{-r/h}.
\end{equation}
This shows that the field vanishes at the origin ($E \sim r^3$), which is 
consistent with the regularity of the geometry and the existence of a de Sitter core. Thus, the reconstructed electric field $E(r)$ follows a qualitatively similar behavior to the Gaussian case presented in Fig. \ref{elefieldgaussian}. While we omit a dedicated plot for this case for the sake of brevity, it is important to note that the $n=1$ configuration exhibits a broader spatial distribution and a slower asymptotic decay ($e^{-r/h}$) compared to the $n=1/2$ case. This reflects the less compact nature of the exponential halo, where the NED sector remains non-negligible over larger radial distances before eventually vanishing.

The detailed analysis of the energy conditions within the electric sector confirms that the regularization of the geometry is a direct consequence of the transverse pressure dynamics. According to the reconstruction formalism, the radial NEC is globally saturated, $\rho + p_r = 0$. Simultaneously, the tangential NEC, expressed as $\rho + p_t = \rho r / (2h)$, remains strictly non-negative throughout the domain, vanishing only at the origin. However, the Strong Energy Condition (SEC), defined also by the combination $\rho + p_r + 2p_t$, effectively reduces to $2p_t = 2\rho (r/2h - 1)$. At the origin ($r=0$), this expression attains the value $-2\rho_0$, which, as in the Gaussian case, reveals a central violation that provides the necessary gravitational repulsion to evade the singularity. As the radial coordinate exceeds the critical value $r=2h$, the SEC becomes positive, demonstrating that the exotic nature of the source is confined to the high-curvature region. This ensures that the electric sector is the sole physical mechanism responsible for the stability of the core and the resolution of the central singularity.

\subsection{Regularity and curvature invariants}

An essential property of the geometries generated by halo density profiles is their regularity. This can be verified by analyzing curvature invariants constructed from the metric function $f(r)$.

For a static spherically symmetric metric gven by Eq.~\eqref{eq1}
the Ricci scalar is given by
\begin{equation}
R = -f''(r) - \frac{4}{r}f'(r) + \frac{2}{r^2}(1 - f(r)).
\end{equation}

Using the relation $f(r) = 1 - 2m(r)/r$, one finds that near the origin,
\begin{equation}
m(r) \sim r^3 \quad \Rightarrow \quad f(r) \approx 1 - \Lambda r^2,
\end{equation}
with $\Lambda = \frac{2}{3}\rho_0$.

Substituting into the curvature invariants, we obtain
\begin{equation}
R \approx 4\Lambda, \qquad R_{\mu\nu}R^{\mu\nu} \approx 4\Lambda^2, \qquad R_{\mu\nu\rho\sigma}R^{\mu\nu\rho\sigma} \approx \frac{8}{3}\Lambda^2,
\end{equation}
which are all finite at $r=0$.

This demonstrates that halo profiles with finite central density naturally generate regular black hole geometries, avoiding curvature singularities. Furthermore, the smooth behavior of the reconstructed electric fields, such as $E(r)\sim r^3$ or $E(r)\sim r^4$, is fully consistent with the regular character of the matter sector.

\section{Dehnen halo geometries}\label{sec6}

Another class of halo models considered in \cite{Konoplya:2025ect} is the Dehnen family

\begin{equation}
\rho(r)=
\rho_0
\left(\frac{r}{a}\right)^{-\alpha}
\left(
1+\frac{r^k}{a^k}
\right)^{-(\gamma-\alpha)/k}.
\end{equation}

This family provides a very general description of halo density profiles and includes several models commonly used in astrophysics.

\subsection{Cases studied in \cite{Konoplya:2025ect}}

For $k=1$ the metric function obtained in \cite{Konoplya:2025ect} takes the form

\begin{equation}
\begin{aligned}
f(r) &= 1-\frac{2m(r)}{r} \\
     &= 1-\frac{2M}{r}
+\frac{2M}{r}
\left(\frac{a}{a+r}\right)^{\gamma-3} \\
&\quad \times
\left[
1+\frac{(\gamma-1)(2a+\gamma r)r}{2(a+r)^2}
\right].
\end{aligned}
\end{equation}

where

\begin{equation}
M=
\frac{8\pi\rho_0 a^3}{(\gamma-3)(\gamma-2)(\gamma-1)}.
\end{equation}

A particularly simple configuration arises for

\[
\gamma=4 ,
\]

for which the metric simplifies to

\begin{equation}
f(r)=
1-
\frac{2M r^2}{(r+a)^3}.
\end{equation}

Near the origin one finds

\begin{equation}
m(r)\sim r^3 ,
\end{equation}

leading to a de Sitter core,

\begin{equation}
f(r)\approx 1-\Lambda r^2 .
\end{equation}

Another family discussed in \cite{Konoplya:2025ect} corresponds to

\begin{equation}
f(r)=
1-
\frac{2M}{r}
+
\frac{2M}{r}
\left(
1+\frac{r^3}{a^3}
\right)^{1-\gamma/3}.
\end{equation}

with

\begin{equation}
M=
\frac{4\pi\rho_0 a^3}{\gamma-3}.
\end{equation}

These geometries interpolate between Schwarzschild behavior at large distances and regular cores at small radii.

\subsection{Regular halo configurations}

When $\alpha=0$ the density remains finite at the origin,

\begin{equation}
\rho(0)=\rho_0 ,
\end{equation}

which implies

\begin{equation}
m(r)\sim r^3 .
\end{equation}

Consequently the spacetime develops a de Sitter core,

\begin{equation}
f(r)\approx 1-\Lambda r^2 ,
\end{equation}

which is a typical feature of regular black-hole geometries.

\subsection{Relation with known regular black-hole solutions}

It is worth noting that particular choices of the Dehnen parameters reproduce geometries already known in the literature of regular black holes. In particular, the case $\gamma=4$ with $k=2$ discussed in \cite{Konoplya:2025ect} corresponds to a regular black-hole geometry which can be obtained within nonlinear electrodynamics minimally coupled to gravity.

Such solutions belong to the same general class as the well-known Bardeen \cite{Bardeen1968}, Hayward \cite{Hayward2006} and Ayón-Beato-García black holes \cite{AyonBeatoGarcia1998,AyonBeatoGarcia1999}, which also possess de Sitter cores characterized by

\begin{equation}
m(r)\sim r^3 ,
\qquad
f(r)\approx 1-\Lambda r^2 ,
\end{equation}

near the origin.

\subsection{Relation with other halo models}

Several astrophysical halo models share the same qualitative structure of a finite central density. A well-known example is the Burkert profile \cite{Burkert1995}

\begin{equation}\label{eq71}
\rho(r)=
\frac{\rho_0}{(1+r/a)(1+r^2/a^2)} .
\end{equation}

This density remains finite at the origin and decreases as $r^{-3}$ at large distances. The corresponding geometry therefore possesses a regular core.

Interestingly, the nonlinear electrodynamics Lagrangian reconstructed from this density exhibits a structure reminiscent of Born--Infeld type theories, suggesting a possible interpretation in terms of effective nonlinear electromagnetic vacuum corrections.

These results indicate that several astrophysical halo profiles share the same geometric structure as regular black-hole solutions generated by nonlinear electrodynamics. In particular, halo models with finite central density naturally lead to spacetimes with de Sitter cores, a feature commonly encountered in regular black-hole geometries.

\subsection{Nonlinear electrodynamics sources}

For the magnetic sector the NED Lagrangian can be obtained explicitly for the halo profiles discussed above from expressions \eqref{eq6}  and \eqref{eq9}.

\paragraph{Burkert profile.}

For the Burkert density \eqref{eq71} the corresponding NED Lagrangian becomes

\begin{equation}
L(F)=
-\frac{\rho_0}{
\left(1+\frac{1}{a}\left(\frac{g^2}{2F}\right)^{1/4}\right)
\left(1+\frac{1}{a^2}\left(\frac{g^2}{2F}\right)^{1/2}\right)
}.
\end{equation}

The electric field $E(r)$ can be obtained by integrating the Burkert density profile, with the mass function $m(r)$ given by
\begin{equation}
m(r) = 2\pi \rho_0 a^3 \left\{ \ln \left[ \left( 1 + \frac{r}{a} \right) \sqrt{1 + \frac{r^2}{a^2}} \right] - \arctan \left( \frac{r}{a} \right) \right\}.
\end{equation}
Now we apply the reconstruction condition from Eq. (\ref{elefieldreconst}). Calculating the required derivatives of $m(r)$ and substituting them into the field equation, we derive the following analytical expression for the electric field:

\begin{equation}
E(r) = -\frac{4\pi \rho_0 r^3}{qa} \frac{\left( 1 + \frac{2r}{a} + \frac{3r^2}{a^2} \right)}{\left( 1 + \frac{r}{a} \right)^2 \left( 1 + \frac{r^2}{a^2} \right)^2}.
\end{equation}

\paragraph{Pseudo-isothermal halo.}

For the pseudo-isothermal density profile

\begin{equation}
\rho(r)=
\frac{\rho_0}{1+r^2/a^2},
\end{equation}

one obtains

\begin{equation}
L(F)=
-\frac{\rho_0}{
1+\frac{1}{a^2}\left(\frac{g^2}{2F}\right)^{1/2}
}.
\end{equation}

To determine the electric field associated with the pseudo-isothermal halo, we need the mass function given by:
\begin{equation}
m(r) = 4\pi \rho_0 a^2 \left[ r - a \arctan\left(\frac{r}{a}\right) \right].
\end{equation}
By substituting the derivatives of $m(r)$ into the reconstruction equation Eq. (\ref{elefieldreconst}), we obtain the analytical form for the nonlinear electric field:
\begin{equation}
E(r) = -\frac{8\pi \rho_{0} a^{2} r^{4}}{q (a^{2} + r^{2})^{2}}.
\end{equation}
In this configuration, the field vanishes at the origin as $r^4$, ensuring a remarkably smooth transition to the de Sitter core. Asymptotically, the field approaches a constant value in the limit of an untruncated halo.

\begin{figure}[h!]
    \centering
    \includegraphics[width=0.9\linewidth]{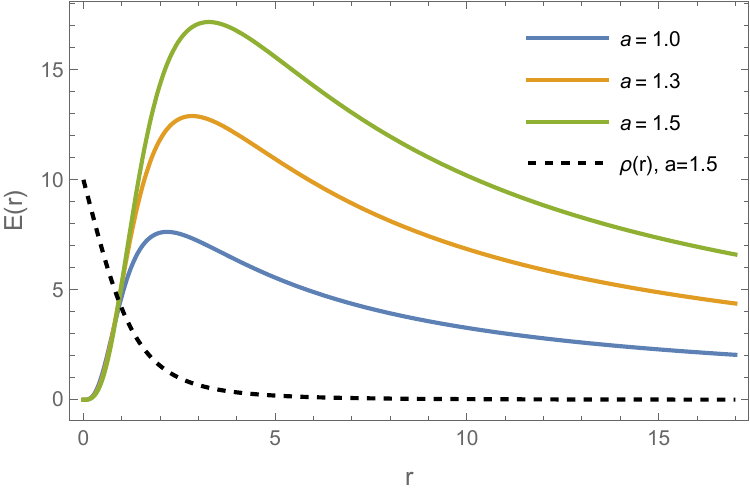}
   \includegraphics[width=0.9\linewidth]{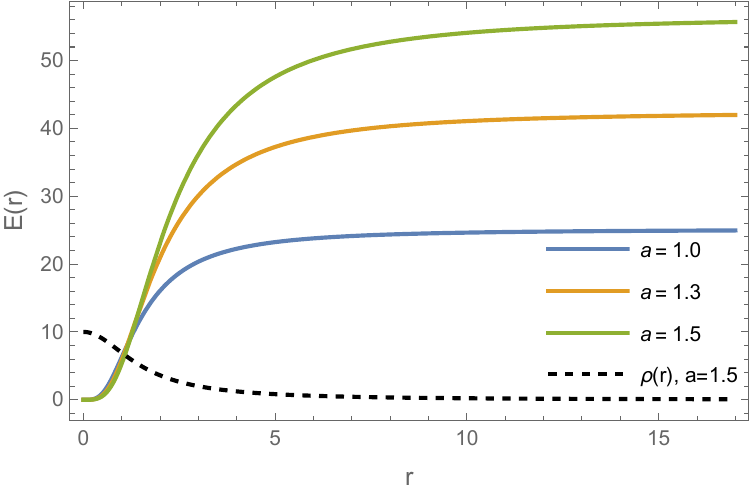}
\caption{Radial profiles of the electric field $E(r)$ and density $\rho(r)$ for various scale radii $a$, with $q = -10$ and $\rho_0 = 10$. The upper panel shows the Burkert profile, featuring an asymptotic decay ($E \sim r^{-1}$). The lower panel displays the Pseudo-Isothermal profile, where the electric field saturates at large distances. In both cases, the black dashed line represents the density for a fixed $a = 1.5$ .}\label{elefieldburkpseudo}
\end{figure}

\paragraph{General behavior.}

For halo models with finite central density the Lagrangian approaches a constant value for large electromagnetic invariant,

\begin{equation}
L(F)\rightarrow -\rho_0
\qquad (F\rightarrow\infty),
\end{equation}

which corresponds to an effective cosmological constant in the central region. At small invariant the Lagrangian typically exhibits inverse power-law corrections,

\begin{equation}
L(F)\sim
-\rho_0
+
\mathcal{O}\!\left(F^{-1/2}\right).
\end{equation}

Such behavior is reminiscent of nonlinear electrodynamics models of Born-Infeld type \cite{BornInfeld1934}, suggesting that halo-supported regular black hole geometries may admit an effective description in terms of nonlinear electromagnetic vacuum corrections.

\section{Universal UV/IR correspondence between halo profiles and NED}\label{sec7}

One of the central results of the reconstruction framework developed in this work is the existence of a universal asymptotic correspondence between the functional behavior of dark matter halo profiles and the ultraviolet and infrared structure of the associated NED theory. This correspondence goes beyond the reconstruction of particular solutions and provides a direct classification scheme relating astrophysical density distributions to universality classes of effective nonlinear field theories.

The key observation is that the magnetic reconstruction establishes a direct relation between the radial dependence of the halo profile and the invariant dependence of the nonlinear Lagrangian. Since the electromagnetic invariant scales as
\begin{equation}
F \sim r^{-4},
\end{equation}
the asymptotic structure of the density profile completely determines the asymptotic behavior of the effective NED sector.

\subsection{Ultraviolet regime and de Sitter universality}

Consider first the near-origin regime. For any admissible halo profile satisfying
\begin{equation}
\rho(r)\rightarrow \rho_0,
\qquad r\rightarrow0,
\end{equation}
with finite central density $\rho_0$, the corresponding mass function behaves as
\begin{equation}
m(r)\sim \frac{4\pi}{3}\rho_0 r^3.
\end{equation}
Consequently, the metric function approaches
\begin{equation}
f(r)\approx 1-\frac{8\pi}{3}\rho_0 r^2,
\end{equation}
which is precisely the de Sitter geometry.

At the level of the NED reconstruction, finite central density implies that the Lagrangian saturates at large invariant,
\begin{equation}
L(F)\rightarrow -\rho_0,
\qquad F\rightarrow\infty.
\end{equation}

Therefore, independently of the detailed functional form of the halo profile, all admissible configurations with finite central density generate the same ultraviolet geometric structure: a de Sitter core supported by an effectively constant NED Lagrangian.

This result reveals that the emergence of de Sitter cores in halo-supported regular geometries is not a model-dependent property, but rather a universal consequence of finite central density. In this sense, regularization occurs through an effective vacuum-like phase in the strong-field regime.

\subsection{Infrared regime and asymptotic field structure}

The large-distance behavior of the geometry is determined by the asymptotic decay of the halo profile. For profiles satisfying
\begin{equation}
\rho(r)\sim r^{-n},
\qquad r\rightarrow\infty,
\end{equation}
the relation between the invariant and the radial coordinate implies
\begin{equation}
L(F)\sim F^{n/4},
\qquad F\rightarrow0.
\end{equation}

This establishes a direct mapping between the infrared behavior of the NED theory and the asymptotic structure of the halo distribution. In particular, the decay exponent $n$ uniquely determines the weak-field scaling of the effective Lagrangian.

Different halo families therefore define distinct infrared universality classes:

\begin{center}
\begin{tabular}{c|c}
Halo asymptotics & Infrared field behavior \\
\hline
$\rho(r)\sim r^{-2}$ & $L(F)\sim F^{1/2}$ \\
$\rho(r)\sim r^{-3}$ & $L(F)\sim F^{3/4}$ \\
$\rho(r)\sim r^{-4}$ & $L(F)\sim F$ \\
\end{tabular}
\end{center}

An important consequence of this classification is that asymptotically Schwarzschild geometries correspond to nonlinear electrodynamics theories approaching the Maxwell limit in the weak-field regime. More general halo profiles instead generate non-Maxwellian infrared sectors characterized by fractional powers of the electromagnetic invariant.

\subsection{Interpretation and implications}

The previous results show that halo-supported geometries encode nontrivial information about the asymptotic structure of the underlying nonlinear field theory. In particular, the ultraviolet sector is controlled by the existence of a finite-density core, while the infrared behavior is determined entirely by the asymptotic decay of the halo profile.

This correspondence provides a direct bridge between astrophysical halo phenomenology and effective NED. Rather than representing arbitrary phenomenological inputs, halo profiles become associated with specific classes of nonlinear field theories characterized by distinct strong- and weak-field limits.

From this perspective, the regularity of halo-supported black hole geometries emerges as a geometric manifestation of nonlinear vacuum effects in the high-curvature regime. The reconstruction framework therefore suggests that regular black hole spacetimes sourced by dark matter halos may admit an effective interpretation in terms of nonlinear electromagnetic vacuum structures, whose ultraviolet and infrared properties are encoded directly in the halo profile itself.

\section{Conclusions}\label{sec8}

In this work we developed a general reconstruction framework establishing a direct connection between dark matter halo profiles and NED minimally coupled to gravity. Starting from static, spherically symmetric halo-supported geometries, we reconstructed the corresponding effective field sources and showed that a broad class of regular black hole spacetimes generated by dark matter halos admits a consistent NED interpretation.

A central result of the analysis is that the reconstruction problem can be formulated directly at the level of the halo density profile $\rho(r)$. In the magnetic sector, the NED Lagrangian is obtained explicitly from the density distribution, while in the electric sector the reconstruction is determined parametrically, revealing the intrinsic non-uniqueness of the inverse problem. This demonstrates that the geometry does not uniquely determine the underlying field theory, but rather defines equivalence classes of NED sectors sharing the same asymptotic gravitational behavior.

Beyond the reconstruction itself, we derived general admissibility conditions ensuring the physical consistency of the resulting configurations. In particular, we showed that halo profiles with finite central density universally generate de Sitter cores and therefore regularize the central geometry independently of the detailed microscopic structure of the NED theory. Conversely, the asymptotic decay of the halo profile determines the infrared scaling behavior of the effective field sector. This establishes a direct correspondence between astrophysical halo phenomenology and the ultraviolet and infrared structure of NED theories.

We applied the formalism to the halo models discussed in~\cite{Konoplya:2025ect}, including the Einasto and Dehnen families, as well as additional cored distributions such as the Burkert and pseudo-isothermal profiles. In all investigated cases, the resulting geometries exhibit regular de Sitter cores and asymptotically Schwarzschild behavior, showing that regularity emerges as a robust and universal feature of halo-supported configurations with finite central density.

An important outcome of the present analysis is that the emergence of regular black hole geometries does not require the introduction of ad hoc modifications of gravity or manually prescribed regular cores. Instead, regularization appears naturally as a consequence of the effective nonlinear structure of the reconstructed matter sector associated with the halo profile itself.

The framework developed here opens several possible directions for future investigation. An immediate extension consists in studying the dynamical and observational properties of the reconstructed geometries, including geodesic structure, quasinormal modes, shadows, and gravitational lensing signatures. It would also be interesting to investigate whether the admissibility conditions derived in this work can be generalized to rotating configurations or to modified theories of gravity. Another important open problem concerns the physical interpretation of the reconstructed nonlinear electrodynamics sector and its possible relation to effective quantum vacuum corrections in strong gravitational fields.

More broadly, the results suggest that dark matter halo phenomenology may encode nontrivial information about the effective field-theoretic structure underlying regular black-hole spacetimes. In this sense, the reconstruction program developed in this work provides a new perspective linking astrophysical density distributions, nonlinear field theories, and the regularization of spacetime singularities within general relativity.

\section*{Acknowledgements}

G. Alencar and C. R. Muniz acknowledge the financial support provided by the Conselho Nacional de Desenvolvimento Científicoe Tecnológico (CNPq), Fundação Cearense de Apoio ao Desenvolvimento Científico e
Tecnológico (FUNCAP) and Coordena\c c\~{a}o de Aperfei\c coamento de Pessoal de N\'{i}vel Superior - Brasil (CAPES).

\bibliography{ref.bib}
\bibliographystyle{elsarticle-num}

\end{document}